\documentclass[12pt]{iopart}

\usepackage{graphicx}

\def\be{\begin{equation}}
\def\ee{\end{equation}}
\def\bc{\begin{center}}
\def\ec{\end{center}}

\def\<{\langle}
\def\>{\rangle}

\def\tp{{t_{\rm per}}}

\def\xip{{\xi_{\rm per}}}
\def\rp{p}

\def\km{k_{\rm min}}
\def\kM{k_{\rm max}}

\begin{document}

\title{Cluster structure and dynamics in gels and glasses}

\author{R Pastore$^{1,2\dag}$, A de Candia$^{1,3,4}$, A Fierro$^{1}$, M Pica Ciamarra$^{1,5}$ and A Coniglio$^1$}

\address{$^1$CNR-SPIN, Via Cintia, 80126 Napoli, Italy}

\address{$^2$UC Simulation Center, University of Cincinnati, and Procter \& Gamble Co., Cincinnati, Ohio 45219, USA}

\address{$^3$Dipartimento di Fisica ``Ettore Pancini'', Universit\`a di Napoli
  ``Federico II'', \\ Complesso Universitario di Monte Sant'Angelo, Via
  Cintia, 80126 Napoli, Italy}
  
\address{$^4$INFN, Sezione di Napoli,
Complesso Universitario di Monte S. Angelo, Via Cintia, Edificio 6,
I-80126 Naples, Italy}

\address{$^5$Division of Physics and Applied Physics, School of Physical and Mathematical Sciences,
Nanyang Technological University, Singapore 637371, Singapore}
\ead{$\dag$ pastore@na.infn.it}

\begin{abstract}
The dynamical arrest of gels is the consequence of a well defined structural phase transition,
leading to the formation of a spanning cluster of bonded particles.
The dynamical glass transition, instead, is not accompanied by any clear structural signature.
Nevertheless, both transitions are characterized by the emergence of dynamical heterogeneities.
Reviewing recent results from numerical simulations, we discuss the behavior of dynamical heterogeneities in different systems and
show that a clear connection with the structure exists in the case of gels.
The emerging picture may be also relevant for 
the more elusive case of glasses.
We show, as an example, that the relaxation process of a simple glass-forming model can
be related to a reverse percolation transition and discuss further perspective in this direction.\\


\end{abstract}

\clearpage

\section{Introduction}
Glasses and gels are 
disordered solids with an arrested dynamics.
According to a gross classification, glasses are high density materials
 dominated by short range repulsion, hard sphere colloids being a paradigmatic case, 
 whereas attractive interactions are necessary to form the low density and fractal structure characterizing gels. 
Polymeric suspensions represent a typical example of systems, where the
inter-particle interaction can be finely tuned to obtain a glass or a gel. 
In absence of cross--linkers or depletants, 
the system is well described as a hard sphere suspension:
the dynamic slow down is driven by increasing the volume fraction $\phi$, 
with the glass transition occurring at a critical threshold, $\phi_{glass}$.
This transition marks the arrest of the dynamics at all relevant length scales, 
for wave-vectors ranging from $\km = 2\pi/L$ to $\kM = 2\pi/\sigma$, with $L$ and $\sigma$ sizes
of the system and of the particles. 
Surprisingly, this sharp dynamic transition is not accompanied by any clear structural signature~\cite{RevCavagna}. 
Standard pair correlation functions, such as the structure factor or the radial distribution function, 
poorly change across the transition, whereas the possibility to find a signature in more complex structural
quantities remains a major and long standing open issue in condensed matter~\cite{RevRoyall}.

A polymeric suspension can also form a permanent gel, even at density much smaller than $\phi_{glass}$, 
on increasing the cross--linker concentration, for instance, by radiation
as in light induced polymerization processes (e.g., dental filling pastes) or by heat (e.g., cooking).
In that way, the system is driven across a percolation transition, where
a spanning cluster of connected monomers emerges. 
This also leads to a dynamical arrest transition, as the dynamics becomes frozen on the smallest wave-vector $\km$, 
the spanning cluster being unable to diffuse. 
Thus, the dynamical transition of permanent gels originates from a well defined structural phase transition (percolation).

Despite the apparent differences between the glass and the gel transition,
both of them are accompanied by the emergence of Dynamic Heterogeneities (DHs)~\cite{DHbook}, groups of particles, spatially correlated over a time scale of 
the order of the relaxation time. DHs have been suggested to play a role similar to critical fluctuations in critical phenomena, and 
seem very promising to distinguish between competing theories and understanding differences and
universality in the dynamical arrest transition.

In this paper, we review recent results on DHs in numerical models of chemical gels (permanent bonds),
colloidal gels (temporary bonds) and colloidal glasses (no bonds), focusing on the connection with the structure.
We start by discussing the case of chemical gels, where
for $k\rightarrow 0$ and $t\rightarrow \infty$, the dynamic susceptibility $\chi_4(k,t)$ tends to the mean cluster size, 
which diverges at the gelation threshold~\cite{tiziana_prl}. 
Then, we consider to what extent this scenario holds on moving
from chemical to colloidal gels~\cite{jstat}, and 
from colloidal gels to colloidal glasses~\cite{noika}, i.e. 
as the slow dynamics is no longer due to the presence of bonds, but rather to particle crowding.
While the framework of chemical gels cannot be trivially extended to glasses, 
it can still lead to novel approaches if the relevant clusters are suitably defined.
As an example, we consider the case of a simple colloidal glass model,
where the relaxation process and the emergence of DHs can be related
to a reverse percolation transition.
As a final perspective, we discuss which percolation model
may describe the glass transition.

\section{Dynamical Heterogeneities in Chemical Gels} 
The transition from solutions (sol) to chemical gels is due to the onset of a permanent spanning cluster,  
which gives rise to the divergence of the viscosity as
the transition is approached from the sol phase, and to an elastic modulus vanishing as the transition is approached from the gel
phase. Since the pioneering work of Flory~\cite{flory, deGennes}, chemical gelation has been explained in terms of percolation models
(for a review see~\cite{stauffer}). 

Here, we consider a model for chemical gels extensively studied in Ref.~\cite{tiziana_prl,tiziana_pre} using Molecular Dynamics (MD) simulations.
Briefly, after equilibrating a system of $N$ particles, interacting via the Weeks-Chandler-Andersen (WCA) potential~\cite{chandler},
permanent bonds between particles at a distance smaller than $1.5\sigma$ 
are introduced, by adding a Finitely Extendable Nonlinear Elastic potential (FENE)~\cite{FENEdum, FENE}.
The numerical simulations have shown a percolation transition at volume fraction $\phi_c\simeq 0.1$, with critical exponents in agreement with random percolation. 
For further details on the MD simulations see Refs.~\cite{tiziana_prl,tiziana_pre}.  
Alternatives models have been described in Refs. \cite{delgado2,saika-voivod}, 
and more recently, in Refs.~\cite{berthier_fene,nagi}.

The emergence of DHs is characterized by  the dynamical susceptibility, $\chi_4(k,t)$ defined as the
fluctuations of the self Intermediate Scattering Function (ISF), $\Phi_s(k,t)$: 
$\chi_4(k,t)=N\left[\rule{0pt}{10pt}\langle
|\Phi_s(k,t)|^2\rangle-\langle \Phi_s(k,t)\rangle^2\right]$,
where $\langle \dots
\rangle$ is the thermal average for a fixed bond configuration,
$[\dots]$ is the average over the bond configurations,
and $\Phi_s(k,t)= \frac{1}{N}\sum_{i=1}^N e^{i\vec{k}\cdot(\vec{r}_i(t)- \vec{r}_i(0))}$.
The wave-vector $k$ allows to probe DHs at different length scales. 

\begin{figure}
\begin{center}
\includegraphics*[scale=0.4]{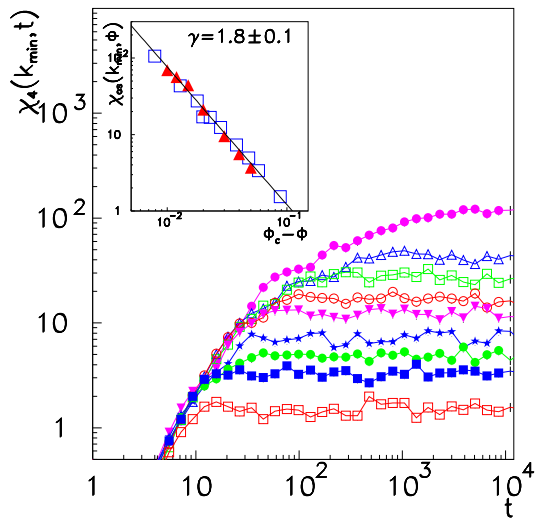}
\caption{
{\bf Main frame}: $\chi_4(k_{min},t)$ as a function of $t$ for
$\phi= 0.02$, $0.05$, $0.06$, $0.07$, $0.08$, $0.085$,
$0.09$, $0.095$, $0.10$ (from bottom to top). {\bf Inset}:
Asymptotic values of the susceptibility (full triangles),
$\chi_{as}(k_{min},\phi)$ and mean cluster size (open
squares) as a function of $(\phi_c-\phi)$.
The data are fitted by the power law
$(\phi_c-\phi)^{-\gamma}$ with $\gamma=1.8\pm 0.1$.
Adapted from Ref.~\cite{tiziana_prl}} \label{fig:1}
\end{center}
\end{figure}

In the main frame of Fig.~\ref{fig:1}, $\chi_4(k,t)$ is plotted as a function of time for $k_{min}=2\pi/L$ and different $\phi$.
In this case, for each value of the volume fraction,  $\chi_4(k_{min},t)$ reaches a plateau after a characteristic time of the order of the relaxation time. 
It was theoretically shown \cite{tiziana_prl}, and numerically verified, that, in the limit of small $k$, the asymptotic value of $\chi_4(k,t)$ coincides with the
mean cluster size (see Inset of Fig.\ref{fig:1}), defined as $S= \sum s^2 n(s) / \sum s n(s)$, where $n(s)$ is the number of cluster of size $s$, which diverges at the gelation
threshold with the random percolation exponent $\gamma$.  
Thus, in chemical gels, DHs are due to the presence of clusters of bonded particles. 
This result also indicates that the percolation exponents can be measured by means
of the asymptotic dynamic susceptibility, and that the asymptotic value of the
dynamical susceptibility plays the same role as the static scattering function near a liquid-gas critical point. 
We note that the dynamical susceptibility shown in Fig.~\ref{fig:1} is qualitatively different from that observed in glasses.
In glasses $\chi_4$ does not asymptotically reaches a plateau, but has a maximum and then decays to zero~\cite{DHbook}.

\section{Dynamical Heterogeneities in Colloidal Gels}
In chemical gels, where the structural arrest is
related to the formation of clusters of bonded particles, the
dynamical susceptibility can be directly connected to the clusters.
This clarifies the nature of the slow dynamics/structural arrest
in these gels as compared to hard sphere glasses~\cite{pusey}, where bonds do not exist at all. 
Very intriguing is the case of gels formed in attractive
colloids~\cite{review_zac}, which are in between these two extremes. 
In these systems, upon tuning the strength of the attractive interactions, one can go
from an irreversible gel, very similar to the chemical gel just
described, to a non-permanent colloidal gel and, finally, to a hard
sphere glass. For example, the addition of non-adsorbing polymers to the suspension induces effective attractive interactions between the colloidal particles due to depletion~\cite{depletion}. In this case, in the temperature-volume fraction plane, the structural arrest line, i.e. where jamming transition occurs, typically
interferes with the coexistence curve~\cite{foffi}.  Gelation may occur, due to an interrupted phase separation, which has been accurately studied in a combined
experimental and numerical effort~\cite{lu_nature}.
The presence of a long range repulsion between particles, due for example to the presence of residual charges, may suppress phase
separation and avoid its interference with gelation. A DLVO-type potential can be used to model this kind of effective interactions,
as often seen in the literature~\cite{noi_dlvo,scior}. Actually, it was shown that the competing attraction and repulsion 
favors ordered columnar and lamellar phases at low temperatures~\cite{noi_tubi,marco}, therefore limiting the possibility to study
the metastable states associated to slow dynamics and structural arrest.  However, upon adding a small degree of polydispersity, it
is possible to avoid the ordered phases and study the arrested line without such interference~\cite{jstat}. 

Since the study of DHs in these systems might
unveil new behaviors arising in intermediate situations and shed new
light on the nature of the structural arrest, 
we now review some findings on a DLVO-type model from Ref.\cite{jstat} and refer
to this paper for further details on the MD simulations.

\begin{figure}[ht]
\begin{center}
a)\includegraphics*[scale=0.37]{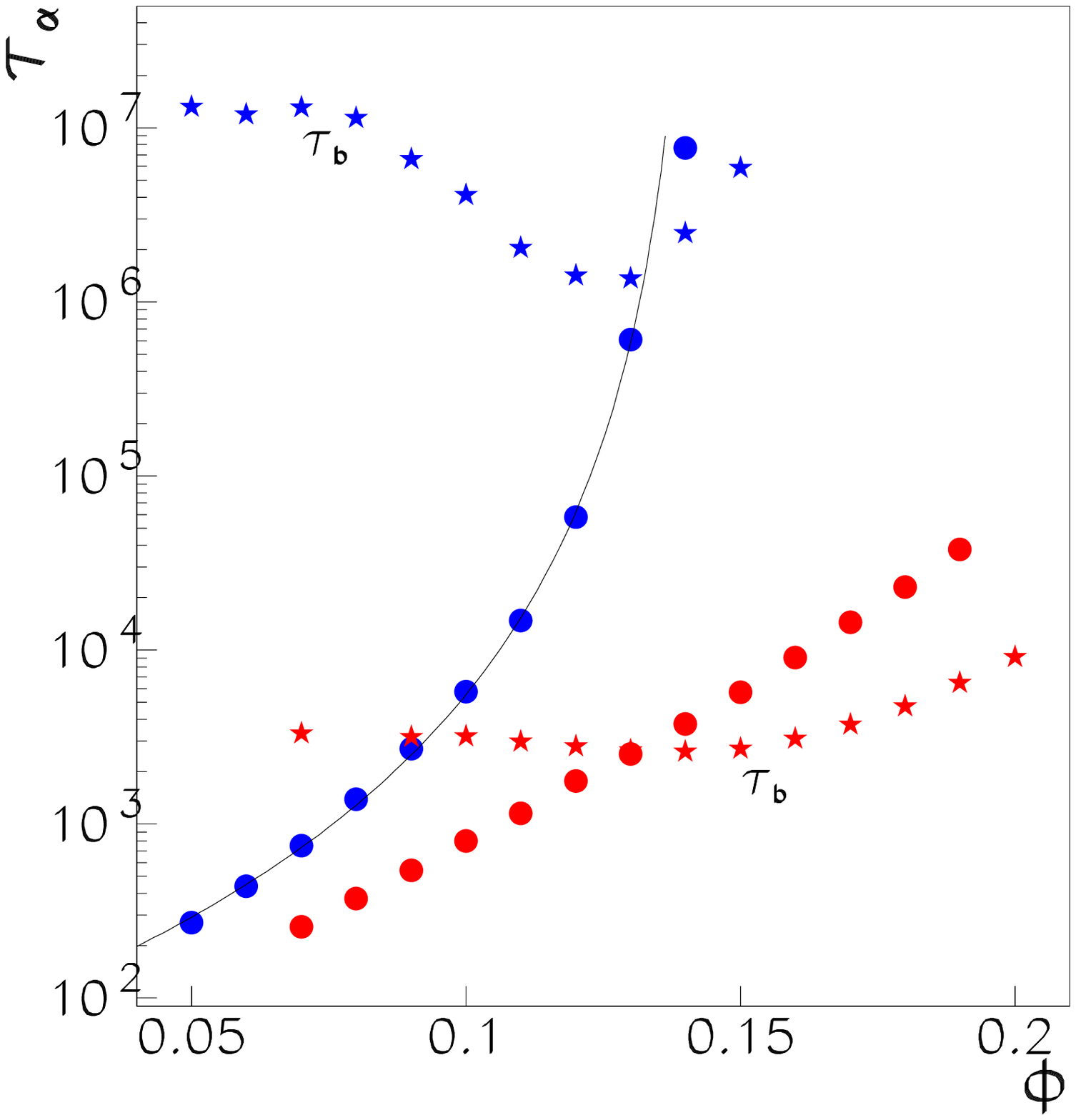}
b)\includegraphics*[scale=0.37]{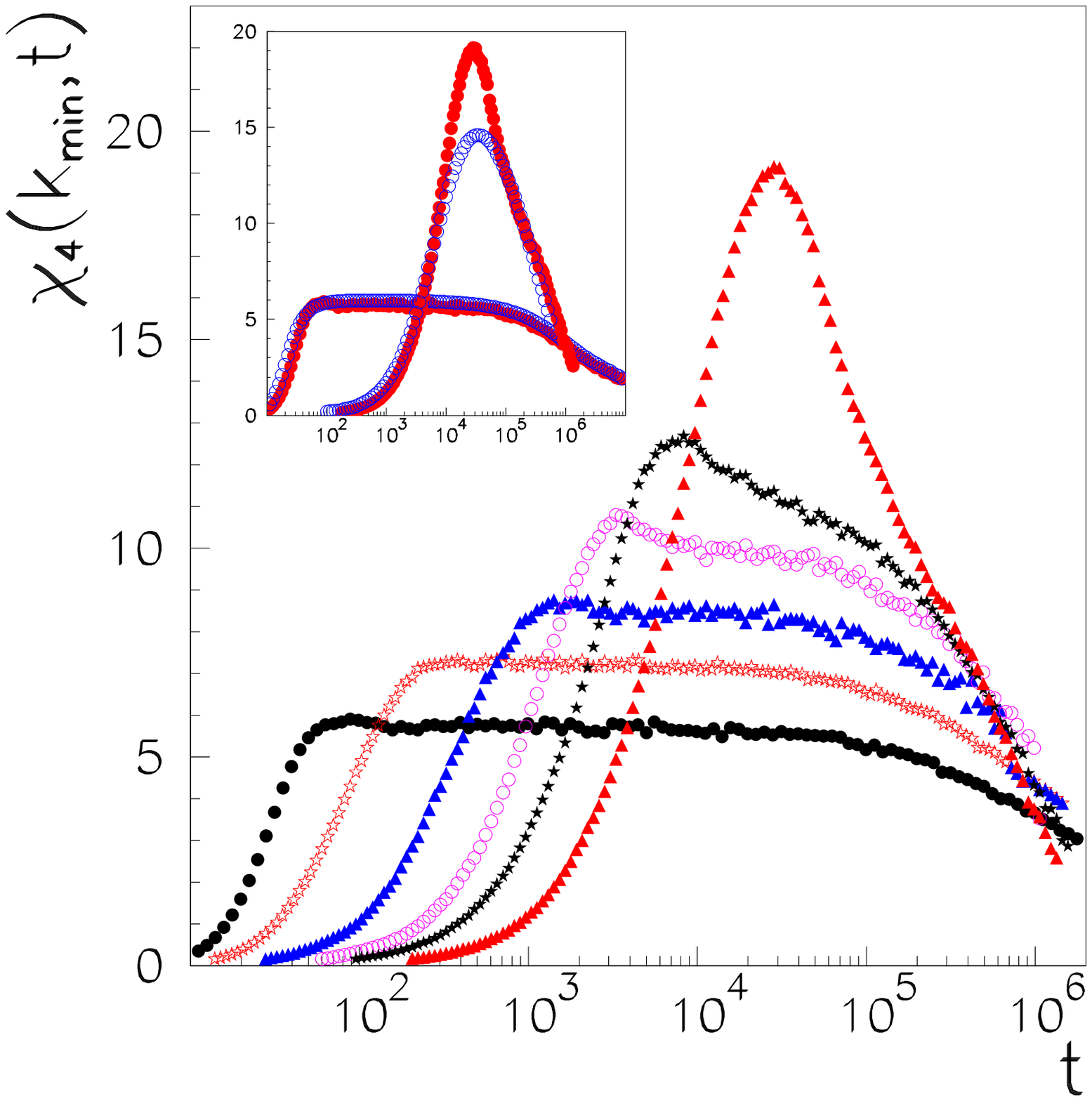}
\end{center}
\caption{
a) The structural relaxation time, $\tau_\alpha(k_{min})$
(circles), compared with the bond relaxation time, $\tau_b$ (stars),
for $T=0.15$ and $0.25$ (from bottom to top). The continuous line is
a power law fit $(0.14-\phi)^{-3.8}$. Adapted from Ref.~\cite{jstat};
b) {\bf Main frame}: The fluctuations of the self ISF,
$\chi_4(k_{min},t)$, obtained in the DLVO model, for $T=0.15$ and
$\phi=~0.01$, $0.05$, $0.08$, $0.10$, $0.11$, $0.12$ (from left to
right). {\bf Inset}: $\chi_4(k_{min},t)$ (circles) compared with
the time dependent mean cluster size, $S_m(t)$, of mobile particles
(open circles), for $T=~0.15$ and $\phi=~0.01,~0.12$. Adapted from Ref.~\cite{jstat}.}
\label{fig:2}
\vspace{-0.3cm}
\end{figure}

In the low temperature and low volume fraction phase, particles tend to form strong bonds, with a lifetime many magnitude orders 
larger than the structural relaxation time, $\tau_b\gg\tau_\alpha$ (see Fig.\ref{fig:2}a).
In this region, the dynamical susceptibility, $\chi_4(k_{min},t)$ (plotted in Fig.\ref{fig:2}b), reaches a plateau, 
roughly coinciding with the mean cluster size.  After a time of the order of the bond lifetime,
$\chi_4(k_{min},t)$ decreases, due to the breaking of the clusters.
This plateau disappears at higher volume fractions, where the bond lifetime is comparable to the structural relaxation time. 
In this case, $\chi_4(k_{min},t)$ eventually exhibits a maximum, as found in hard sphere glasses.
These results clearly demonstrate that, when the bond lifetime is long enough, as compared to the relaxation time, the behavior of the dynamical susceptibility is akin to
that measured in chemical gels, on time scales for which the bonds can be considered as permanent.
At longer times, the breaking of the clusters causes the final decay to zero.
A geometrical interpretation of $\chi_4(k_{min},t)$ in this system can be given by considering
the time dependent mean cluster size $S_m(t)$ of mobile particles
 \cite{jstat}. Here two particles are considered as part of the same cluster if they are 
bonded both a time zero and at time $t$.
In the inset of Fig.\ref{fig:2}b), $S_m(t)$
is plotted for volume fraction  $\phi=0.05$ and  $\phi=0.12$. At low
volume fraction the coincidence of $\chi_4(k_{min},t)$ and $S_m(t)$
is excellent. At higher volume fraction the maximum of
$\chi_4(k_{min},t)$ is larger than the maximum of $S_m(t)$, denoting
that the contribution to the peak comes not only from the cluster
formation, but also from the crowding of the particles, as usually observed in glassy systems.
The figure exemplifies that there are two different mechanisms underlying the presence of significant DHs
at different volume fractions. It also shows the crossover from the cluster dominated regime to the crowding dominated regime.

\section{Dynamical Heterogeneities in Structural Glasses}
We now discuss how arguments from percolation and gels can be useful to characterize the relaxation process and DHs in 
glasses, where particles interact via a hard core repulsion, and the dynamics slows down at high density due to crowding.
To this aim, we review recent results~\cite{noika, Fractals} from numerical simulations of the KA lattice model~\cite{KA}.

The KA is a kinetically constrained model~\cite{Ritort}, in which a lattice of volume $V$
is randomly occupied by a number $\rho V$ of non
overlapping particles, and each particle is allowed to move in a near empty site 
if has less than $m= 4$ neighbors before and after the move.  Previous studies have shown that this model reproduces many aspects of
glass forming systems, as the dynamics slows down on increasing the density, and suggests the existence of a transition of
structural arrest at $\rho_{ka} = 0.881$~\cite{KA}.
Even though, it has been demonstrated that in the thermodynamic limit the transition of dynamical arrest only occurs at $\rho = 1$~\cite{Toninelli}.

\begin{figure}
\begin{center}
\includegraphics*[scale=0.4]{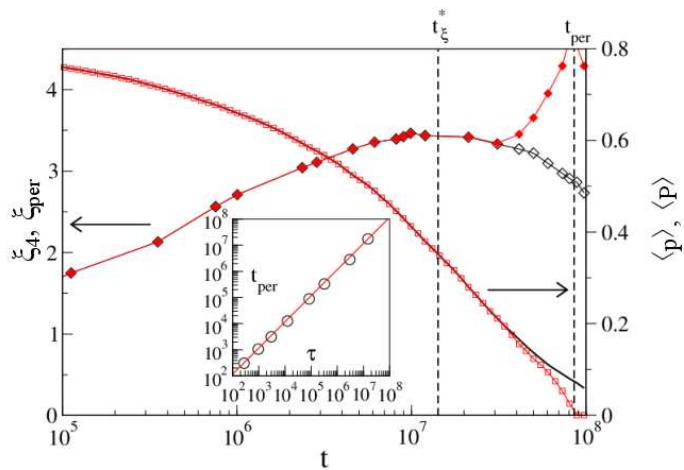}
\caption{ \label{fig:4}
Relaxation process and reverse percolation in the KA model at $\rho = 0.87$.
Left axis: dynamical correlation length $\xi_4$ (empty diamonds) and percolation
correlation
length $\xip$ (full diamonds). Before the two lines start to separate at large time, empty and full diamonds coincide. Right axis: density of persistent particles
$\<\rp\>$ (full line)
and strength of the percolating cluster $\<P\>$ (squares).
The vertical dashed lines mark $t^*_\xi$ and $\tp$, which is proportional
to $\tau$ (inset). Adapted from Ref.~\cite{noika}.
}
\end{center}
\end{figure}

After introducing an occupation number $n_i(t) = 1(0)$, if site $i$ is (is not) persistently occupied by a particle in the time interval $[0,t]$, we
monitor the relaxation process through the  density of persistent particles, $\rp(t)=\frac{1}V \sum_{i=1}^{V} n_{i}(t)$,
which is related to the high wave-vector limit of the self ISF~\cite{Chandler2006}.  
Accordingly, we define the structural relaxation time from the
relation $\<\rp(\tau)\>/\rho=e^{-1}$.  
The decay in time of the ensemble average, $\<\rp(t)\>$, for a high density value is illustrated in Fig.~\ref{fig:4}.

Similarly, we quantify the emergence of DHs focusing on the spatial correlation function between persistent particles at time $t$~\cite{Fractals},
\begin{equation}
\label{eq:gr}
g_4(r,t)=\<n_{i}(t)n_{j}(t)\>-\<n_{i}(t)\>\<n_{j}(t)\>,\,\,\, r = |i-j|.
\end{equation}
and, from its spatial decay, we extract a dynamical correlation length $\xi_4(t)$.
\begin{figure}[t!]
\begin{center}
\includegraphics*[scale=0.9]{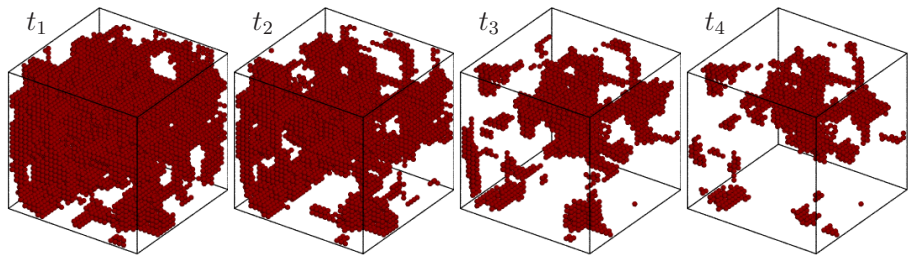}
\end{center}
\caption{\label{fig:3}
(Color online)
Persistent particles in the KA model at $\rho=0.85$ and increasing time from left to right.
$t_3$ coincides with the structural relaxation time. Adapted from Ref.~\cite{noika}
}
\end{figure}

\begin{figure}[ht]
\begin{center}
a)\includegraphics*[scale=0.40]{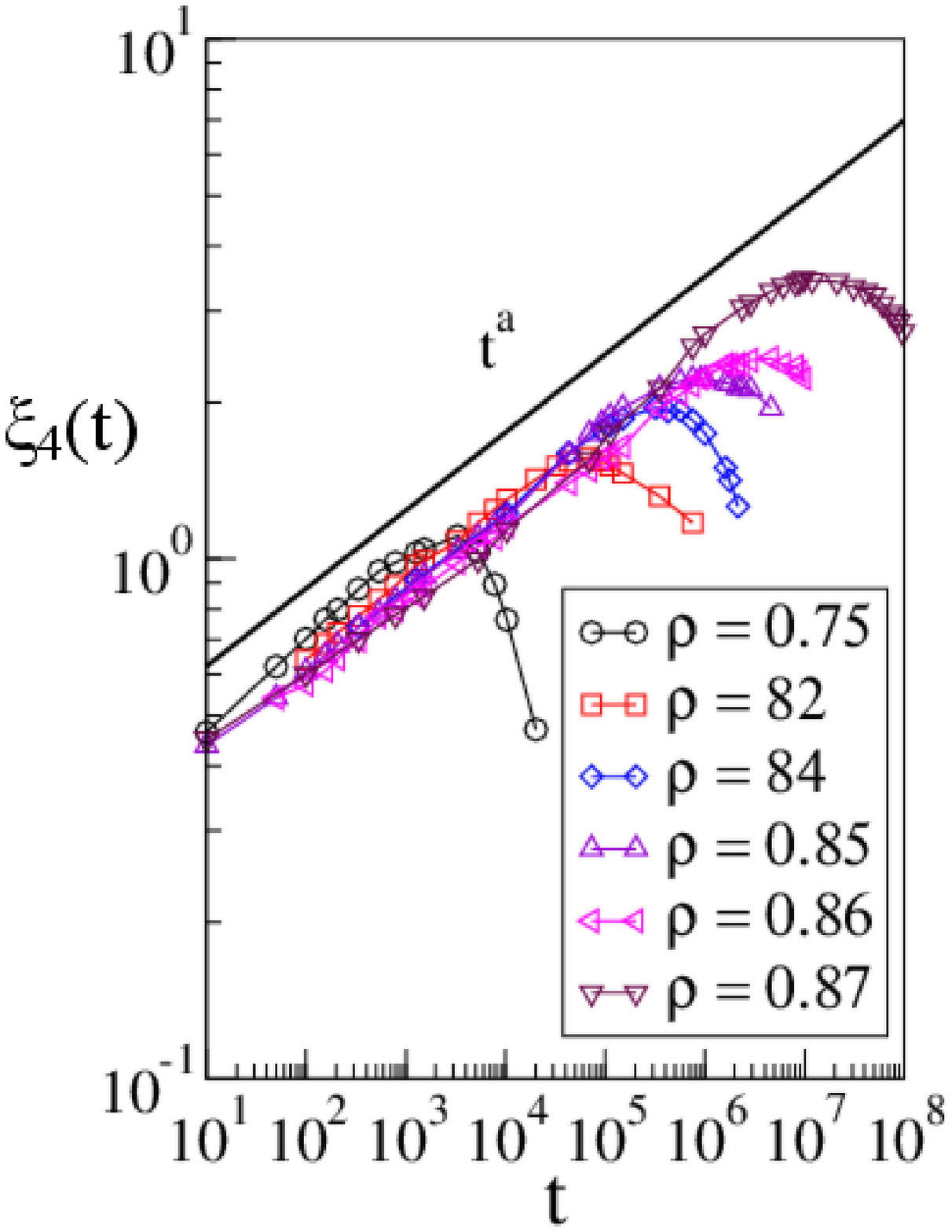}
b)\includegraphics*[scale=0.42]{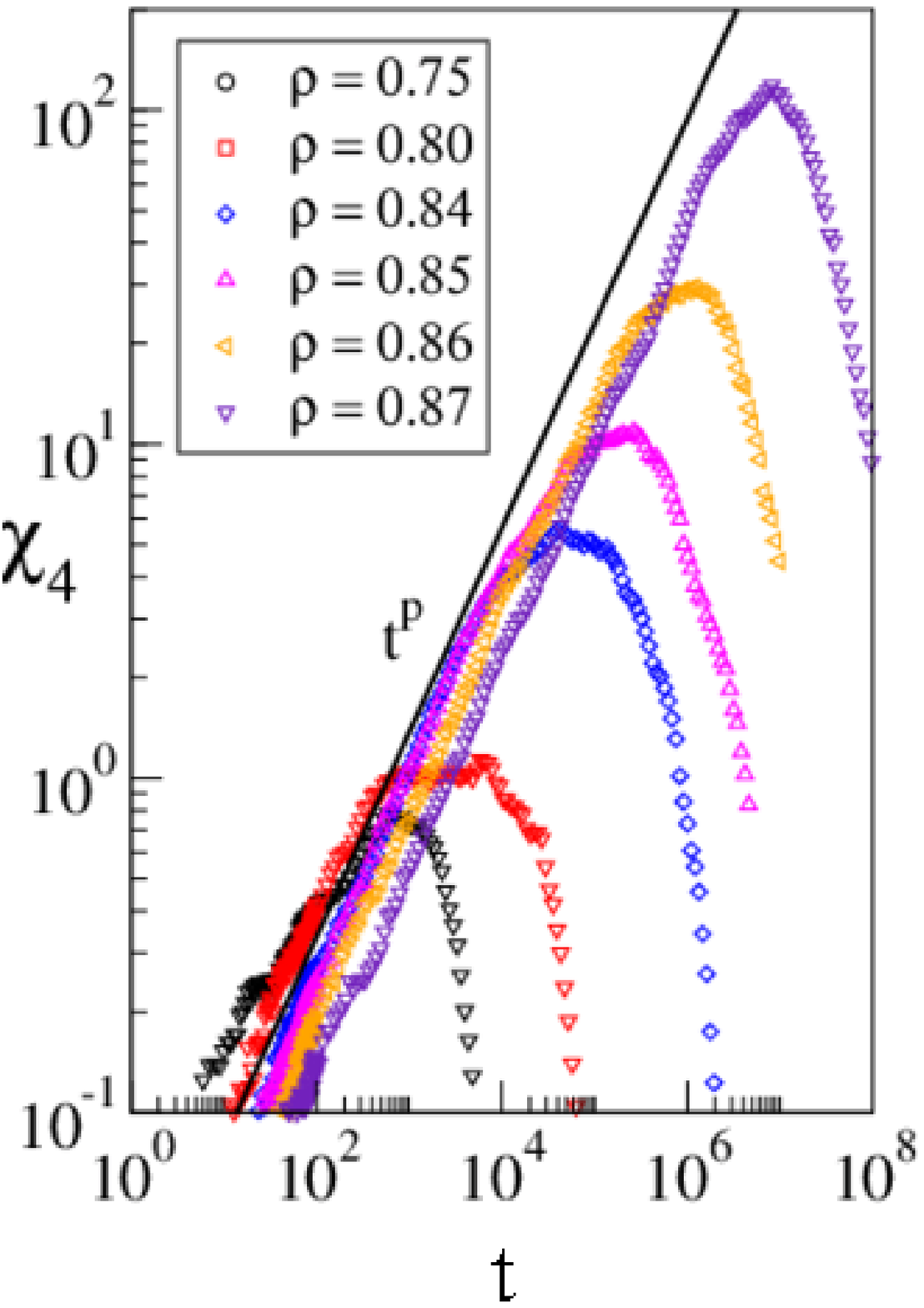}
\end{center}
\caption{
a) Dynamical correlation length $\xi_4(t)$ and b) dynamical susceptibility $\chi_4(t)$ for the KA model at different density, as indicated.
The initial increase is well described by the power laws reported as solid lines, $t^a$ with $a\simeq0.16$ and $t^p$ with $p\simeq0.6$, respectively.
Adapted from Ref.~\cite{noika}
}
\label{fig:3bis}
\vspace{-0.3cm}
\end{figure}

Fig.\ref{fig:3} suggests that, as time andvances and the number of persistent particles decreases, 
spatial correlations first grow and then decay.
Accordingly, Fig.\ref{fig:3bis}a shows that $\xi_4(t)$ exhibits, at short times, a power-law increase, and then decreases after reaching its maximum value $\xi^*$ at time $t^*_\xi$. Both $\xi^*$ and $t^*_\xi$ increase as power laws of $\rho_{ka} -\rho$.
Fig.\ref{fig:3bis}b shows that the dynamical susceptibility, $\chi_4(t)=\frac{V}\rho\left(\<\rp(t)^{2}\>-\<\rp(t)\>^{2}\right)=\frac{1}{\rho V}\int_{V}g_4(r,t)d^3r$
has a behavior qualitatively similar to $\xi_4(t)$~\cite{noika} and allows to appreciate the difference with the gel case, as discussed before.
We also found~\cite{noika} that the relation between the behavior of $\langle p(t)\rangle$, $\xi_4(t)$ and $\chi_4(t)$ 
can be rationalized within the diffusing defect paradigm~\cite{defects1, defects2}, which ascribes the relaxation of the system to
the diffusion of possibly extended defects.

To provide a geometrical characterization of this scenario, we consider that the glass former may be 
thought as rigid on time-scales smaller than the
relaxation time $\tau$, and as liquid at larger times.  
In that sense, we expect that, as long as the system behaves as solid, a percolating cluster 
of persistent particles plays the role of the physical backbone in gels, and that
a reverse dynamical percolation transition occurs for time-scales of
the order of the relaxation time. Indeed, the absence of the percolating cluster should lead to the loss of rigidity.
In order to investigate whether this supposed transition was related to the relaxation process,
we define a dynamical kind of bond, similarly to the case of
colloidal gels: two particles $i$ and $j$ are bonded in the interval $[0,t]$ if they are nearest neighbors and persistent in this time interval.

In Fig.\ref{fig:4}, we show $\langle P(t) \rangle$, the density of persistent particles belonging to the spanning 
cluster. $\langle P(t) \rangle$ vanishes at a time $t_{per}$, which is found to scale with the relaxation times, $\tau$, as the density increases (Fig.\ref{fig:4}, inset).
The figure also reveals that the cluster strength overlaps with the total density of persistent particles, $\langle P(t) \rangle \simeq  \langle p(t)
\rangle$, up to large times.  This means that in this interval the percolating cluster is the only
cluster present.  At larger time, $\<p(t)\>$ slowly decays, while $\<P(t)\>$ vanishes,
because of the emergence of finite clusters, with a broad size distribution, 
which give contributions to $\langle p(t)\rangle$, but not to $\langle P(t)\rangle$.
This circumstance may explain the crossover from power-law to stretched
exponential observed for $\< p(t) \>$ \cite{noika}: 
indeed the short time decay is characterized by a single relaxation time,
whereas the long time decay results from a broad spectra of finite cluster lifetime.

To better understand the geometrical properties of this process, we investigate the correlation length, $\xi_{per}(t)$, which
is defined by the percolative correlation function $g_{per}(r,t)$:
 \begin{equation} \label{eq:g_per}
 g_{per}(r,t)=P_{i,j}^{f}(r,t)+P_{i,j}^{\infty}(r,t) -\langle P(t) \rangle^2
 \end{equation}
where $P_{i,j}^{f}(r,t)$ and $P_{i,j}^{\infty}(r,t)$
are the probabilities that two sites $i$ and $j$, in the configuration of persistent particles at time $t$, belong to the same finite cluster, or to the percolating cluster respectively.
Accordingly, if the percolating cluster is absent, $g_{per}(r,t)=P_{i,j}^{f}(r,t)$, and $\xi_{per}(t)$
measures the typical size of finite clusters. Conversely, if finite clusters are negligible, then
\begin{equation} \label{g_per_inf}
g_{per}(r,t)=P_{i,j}^{\infty}(r,t)-\langle P(t) \rangle^2
\end{equation}
measures the extension of the density fluctuations within the percolating cluster.
In our case, at short time, $\langle P(t) \rangle \simeq  \langle p(t) \rangle$ and $P_{i,j}^{\infty}(r,t)\simeq \langle n_i(t)n_j(t) \rangle_{|i-j|=r}$,
because almost all persistent particles belong to the percolating cluster, making the connectedness condition negligible~\cite{Coniglio}.
Inserting these equalities in Eq.(\ref{g_per_inf}) and comparing it with the definition of the four point correlation function, $g_4(r,t)$, Eq.(\ref{eq:gr}), 
we find that $g_{per}(r,t)\simeq g_{4}(r,t)$, and consequently $\xi_{per}(t)\simeq\xi_4(t)$.
Indeed, Fig.\ref{fig:4} confirms that the dynamical correlation length coincides with the percolative length, as long as finite clusters are negligible.

The fact that finite clusters are essentially absent until a spanning cluster exists,
suggests that this percolation is different from the random percolation, which drives instead the chemical gelation.
Thus, what is the type of percolation relevant to describe the relaxation process in the KA model?
The scenario emerging from the related mean field case could be useful to tackle this issue.  
In a recent paper~\cite{decandia}, the dynamical behavior of the Fredrickson and Andersen (FA) facilitated model \cite{FA}, which is a spin version of the KA model, was indeed studied  on the Bethe Lattice. 
In the infinite time limit, this model reproduces
\cite{sellitto} the Bootstrap Percolation (BP) model~\cite{chalupa,schwarz,baxter2011}. 
In BP, the lattice sites are occupied randomly with density $\rho$. Then each
particle, which has less than a fixed number $m$ of  neighbors, is removed, 
until a stable configuration is reached. 
This configuration is made of clusters of particles, where each particle
has less than $m$ neighbors.
BP on the Bethe lattice of coordination number $z$ has a mixed order transition when  $m>2$.  
Below the percolation threshold, there are no occupied sites, while above the threshold there are 
no finite clusters, but only an infinite cluster which is called $m-$cluster.
The percolation order parameter, $P$, jumps discontinuously at the threshold from $0$ to $P_c$, and 
the approach to $P_c$ from the percolating phase is characterized by 
a critical exponent $\beta=1/2$~\cite{chalupa},
while the fluctuation of the order parameter  and the associated length diverge with  
exponents $\gamma =1$ and $\nu= 1/4$~\cite{decandia}, respectively. 

Summarizing, in the FA model, the infinite time limit of the dynamical order parameter, its  
fluctuation and the dynamical length are given by the corresponding static quantities of BP.
Consequently, these static quantities are different from zero only in the glassy phase, and 
become critical at the threshold. Nevertheless, in the liquid phase,  
the time dependent dynamical quantities $\langle p(t)\rangle$, $\xi_4(t)$ and $\chi_4(t)$ of the FA model, although 
tend to zero in the infinite time limit, reflect at finite time 
the behavior of the BP static quantities~\cite{decandia} .  
The picture is similar to that found in the liquid phase of colloidal gels, in the particular case in 
which  the bond lifetime and the relaxation time become comparable,
with the difference that 
random percolation drives the dynamical transition in gel models, 
whereas bootstrap percolation drives the glass transition in the FA model.
Interestingly enough there is strong evidence \cite{sellitto,ArSe,FrSe,maurocrossover,decandia} that 
the dynamical behavior of the FA model on the Bethe lattice and of gel systems 
in mean field are related~\cite{arenzon} to the ones predicted by the discontinuous and the continuous Mode Coupling Theory (MCT)~\cite{Gotze}, respectively. 


\section{Conclusions}
The study of DHs allows to clarify the nature of slow dynamics and structural arrest in gels and glasses.
The behavior of the dynamical susceptibility in chemical gelation is quite different from that found in 
hard sphere glasses. It grows steadily and reaches a plateau, whose value, in the low wave-vector limit,
coincides with the mean cluster size.  
In colloidal gelation, at low temperature, DHs are associated to clusters made of long living bonds and 
the dynamical susceptibility reaches a plateau, as found in chemical
gels, except that it decays to zero at long times, due to the finite
lifetime of the clusters. 
DHs in physical gels are also analyzed in Ref.~\cite{berthier_fene}, where
a FENE model with finite bond lifetime is studied using molecular dynamics 
simulations. However, the dynamical susceptibility, evaluated only at large wave-vectors, displays the behavior usually found in glassy systems. We expect that also in that case, when
the bond lifetime is long enough, the dynamical susceptibility in the gel
phase displays,  at small wave-vectors, a plateau at intermediate times,
signaling the presence of persistent clusters.
At higher volume fraction, DHs crossover to a different behavior, where crowding
starts to play a role, which becomes dominant for glasses.
In this case, bootstrap percolation model seems to provide a geometrical characterization of DHs.

Finally, we suggest that the use of very recent optical technique, such as
the Digital Differential Microscopy~\cite{CerbinoPRL, GiavazziRev}, might provide new experimental 
insights on DHs of colloidal systems in a wide range of wavelengths.

\section*{Acknowledgments}
We acknowledge financial support from MIUR-FIRB
RBFR081IUK, from the SPIN SEED 2014 project
{\it Charge separation and charge transport in hybrid
solar cells}, and from the CNR-NTU joint laboratory {\it Amorphous materials for energy harvesting
applications}.

\end{document}